\documentclass[twocolumn,aps,floats,showpacs,amsmath]{revtex4}
\usepackage{amssymb,amsmath}
\usepackage{graphicx,float}

\tighten
\begin{document}
\title{ \Large \bf
Traffic on complex networks: Towards understanding global 
statistical properties from microscopic density fluctuations}
\author{\large  \flushleft 
Bosiljka Tadi\'c$^{1}$,  
Stefan Thurner$^{2}$ and G. J. Rodgers$^3$
} 

\affiliation{ 
$^1$Department  for Theoretical Physics; Jo\v{z}ef Stefan Institute, 
P.O. Box 3000; SI-1001 Ljubljana; Slovenia, 
$^2$Complex Systems Research Group HNO AKH; Universit\"at Wien, 
W\"ahringer G\"urtel 18-20; A-1090 Vienna;  Austria,
$^3$Department of Mathematical Sciences; Brunel University, Uxbridge; Middlesex UB8 3PH; UK\\ \hspace{1cm}} 

\begin{abstract}
\noindent
We study the microscopic time fluctuations of traffic-load and the global 
statistical properties of a dense traffic of particles on scale-free cyclic 
graphs. For a wide range of driving  rates $R$ the traffic is stationary 
and the load timeseries exhibit 
anti-persistence due to the regulatory role of  
the super-structure associated with two hub nodes in the network.
We discuss how the super-structure effects the functioning of the 
network at high traffic density and at the jamming threshold. 
The degree of correlations  systematically
decreases with increasing traffic density and eventually 
disappears when approaching a jamming density $R_c$. 
Already before jamming we observe qualitative changes in the 
global network-load distributions and the particle queuing-times.
These changes are related to the occurrence of 
temporary crises in which the network-load increases dramatically, 
and then slowly falls back to a value characterizing free-flow. 
\noindent
\end{abstract}
\maketitle

\section{Introduction}

Microscopic dynamic processes and emergent statistical properties are two 
facets of complex dynamic
systems that are closely connected, and for many systems 
understanding their interdependence
is important for both prediction and 
strategic planning. 
Usually detailed information is available about either 
the microscopic dynamics or the statistical properties, 
but seldom both. 
Prominent examples include
 traffic noise on communication networks 
\cite{traffic_exper1,traffic_exper2}, noisy signals in driven 
disordered and self-organized systems \cite{BHN_minirev}, and the timeseries 
of price fluctuations in financial markets \cite{econophysics}.
On the other hand, emergent behavior in a statistical system 
on a macroscopic scale can be described 
by (stable) statistical laws,
their sensitivity to relevant parameters of the dynamics can be studied 
and the type of global behavior may be predicted.  
In this work we use a recently proposed  model \cite{TT} to study 
both the  microscopic fluctuations of 
traffic timeseries on  complex networks 
and the statistical properties of transport of individual 
particles. These particles can be thought of, e.g., as information 
packets in the Internet or organizations, or proteins transported on 
the cytoskeleton 
of a cell. We study how the statistical properties,  
both at a microscopic and macroscopic levels, vary with  the particle  
creation rate $R$, i.e., traffic density.

A new class of networks, called scale-free, has been recognized as 
the most commonly observed network structure, which appears to be 
also the most stable (see, e.g., \cite{DM_book,AB}).
In particular, communication networks such as the Internet and the Web 
are scale-free networks  with both in-coming
and out-going link connectivity distribution 
obeying a power-law with significant clustering and
link correlations \cite{watts,Web,vazquez}. 
A model graph  with these properties, which we call ''Web-graph'' was 
recently proposed \cite{BT}.
For study of particle traffic on complex networks we recently 
introduced a model of simultaneous random walks on 
scale-free cyclic and tree-graphs \cite{TT,TR}. In low particle density the
distributions of particle transit-times exhibit power-law dependences 
with the exponents depending on the network structure. Other theoretical 
studies capture the essential properties of the jamming transition and
self-tuned driving on simpler topologies, like hierarchical trees \cite{BCN1}
and square lattice models \cite{Sole1,Sole2}.
 
Due to finite processing capacities of nodes traffic queues occur
especially at hub nodes, depending on the intensity of traffic. 
This means that the transit-time 
for particles depends not just on the distance between the sending 
and receiving node but also on the geometry and local traffic density 
\cite{TT}.
The distribution of transit-times is important 
for  network efficiency and for estimating the risk of transport delay.
For a given graph and a search algorithm, a fundamental quantity 
that contributes to the emergent transit-time is the waiting-time 
that a particle spends in queues along its path.

In \cite{TT} we considered different 
network topologies and showed 
how the dilute, sparse, topology of the network influences the 
transport on it. Different search 
algorithms were employed with low density traffic to quantify the 
network's performance. 
We have found that (i) low density traffic is stationary; 
(ii) away from the jamming transition the  
distribution of transit-times is power law due to the topology; 
(iii) the waiting-times are small
for low traffic density;
(iv) the Web-graph topology of a scale 
free directed network with closed loops and a next-nearest-neighbor 
search strategy
 results in efficient traffic with a large output rate
which utilizes the hubs effectively.
Consequently, compared to many other topologies the Web-graph can 
support a huge volume of traffic before getting jammed.

In this paper we report on  a complementary study. 
We consider the Web-graph, and analyze the waiting-time  
statistics and the timeseries of 
network-load, which is defined as the number of particles on 
the network at a given time, as the traffic density is 
varied by increasing the creation or posting rate $R$. 
We also consider correlations  in the timeseries of the 
network's activity, i.e., the number of simultaneously 
active nodes in the network. 
The queuing of particles on different nodes is 
a consequence of 
a self-regulatory traffic of mutually interacting random 
walks sent to different specified destinations on the graph.
Throughout the paper we use the term {\it load} in a functional
sense, as defined above. It should not be confused with a 
topological meaning which is sometimes used for the number of minimal 
paths through a node \cite{kahng} or betweenness \cite{newman}.  

The paper is arranged as follows. In the next Section we describe the
model in detail, in Section 3 we consider the correlation in the 
load timeseries and in Section 4 the distributions of waiting-times and 
network-loads are presented. 	In Section 5 the work is summarized.

\section{Graph structure and traffic rules}

The Web-graph is a directed graph grown with microscopic 
dynamic rules  proposed in \cite{BT}, which were 
originally intended to model the evolution of the 
world-wide Web. It belongs to a class of models with preferential attachment
of nodes \cite{AB,DM_book}.
In addition to preferential attachment of newly added nodes, the  rules 
of Web-graph  evolution include rewiring of preexisting links while 
the graph grows, which results in an emergent
structure with inhomogeneous scale-free ordering in both in-coming 
and out-going links and a number of closed cycles. 
An example of such an emergent structure of a Web-graph is shown in 
Fig.~1. 

\begin{figure}[tb]
\begin{center}
\begin{tabular}{c} 
\includegraphics[width=8.2cm]{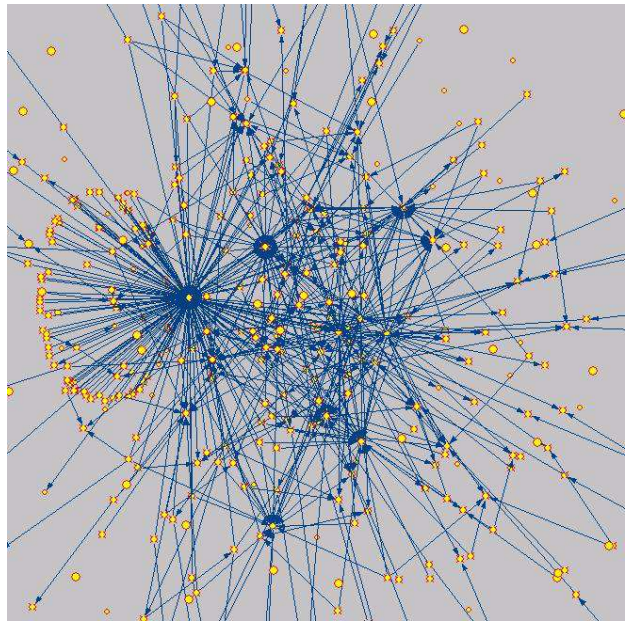}
\end{tabular}
\end{center}
\caption{Web-graph consisting of $N=L=1000$ nodes and links.}
\label{figgraph}
\end{figure} 

A detailed characterization of
the topology of this Web-graph, both on a local and global level, 
together with a  discussion of the   
origin of scaling laws in this system, can be found in 
\cite{BT,BT_ccp02,BT_fract,BT_ccp03,Granada}.
The observed power-law increase of local connectivity in 
time \cite{BT_fract} 
can be linked to the power-law distributions for in- and 
out-going links, each with their own
scaling exponents \cite{BT}. Other topological properties 
which are relevant for
traffic on the graph were also studied, in particular 
the number of paths through a 
given node as a measure of betweenness was found to have a power-law 
decay \cite{Granada} and the
degree of clustering and the correlations between in- and 
out-connectivity \cite{BT_ccp03} resemble the properties 
measured in empirical studies of the 
Web \cite{Web} and the Internet \cite{vazquez}. 

An important feature of the graph topology is the occurrence of 
two types of hub nodes
(known as hub and authority nodes in the real Web). 
These are nodes with large in-coming link
connectivity and nodes with large out-going link connectivity. 
These two hubs have many links between them and impose what we call   
a super-structure 
which has a strong influence on the transport processes on the graph 
\cite{TT}. 

We grow the graph to a given size 
($N=L=1000$ nodes and links) and fix the
connectivity matrix after growing. We then model the traffic of 
particles on that graph \cite{TR,TT}. 
Particles are created with a given rate $R$ (particles per time-interval) 
at  randomly selected nodes and are given a randomly selected destination 
node where it should be delivered. We select 
these pairs of nodes within the giant component of the graph. 
Particles move through the graph simultaneously 
searching for their respective destination addresses. 
To navigate particles, each node 
performs a local search in its next-nearest-neighborhood, 
and if the particle's destination 
is found within the searched area, it is delivered to the 
node's neighbor linked to 
the destination node. Alternatively, the particle moves to a randomly 
selected neighbor.
This search algorithm was shown \cite{TT} to perform especially well 
on the Web-graph,
where it can effectively make use of the hub nodes. 
In particular, it was shown that it 
performs 40 times better than a random diffusion on the same graph and 
8 times better than the
same algorithm on a scale-free tree graph with the same in-link connectivity. 

Additional rules are necessary to regulate the traffic. 
We assign a buffer 
($H$=1000) to each node in the network. When the buffer at 
a selected node is full the node cannot accept more particles
and the particle waits for the next opportunity to be delivered. 
Due to the simultaneous movement of particles queues can be formed, 
especially at 
nodes with large connectivity.
Here we apply the LIFO (last-in-first-out)
queuing discipline at each queue. 
When a particle arrives at its destination it is removed from the network. 
For simplicity, in this work we allow particles to move along out-going links 
and against in-coming links with equal probability. 
Details of the implementation of the numerical code are given in
\cite{BT_ccp03}.

\section{Correlations in network-load timeseries}

Each particle follows its own (random) path from the 
origin to its destination. The total time spent
along the path, the {\it transit-time}, depends on both the topology 
and the time the particle spends waiting in queues along that path. 
Statistics of transit-times on different 
topologies with different search algorithms were studied in \cite{TT}. 
With an efficient search,  transit-times can be short, however, some 
particles get into remote areas of the graph, from which it takes a 
long time to escape; this
results in a power-law distribution of transit-times. 

With increasing posting rates the interaction between particles, 
i.e., queueing in hubs,  becomes 
more important and results in longer queuing-times. 
In effect, the number of particles on the network 
fluctuates in time in a way that
is characteristic of the network structure and the search algorithm, and 
depends on the posting rate and the buffer capacity. 
We find that for a wide range of posting rates $R$ the 
traffic is stationary
with the average network output (number of particles 
delivered per time-step) balancing the input rate $R$. 
The load fluctuates around an average value, which increases with $R$. 
Eventually, for large $R>R_c$ a permanent increase in the number of 
particles occurs, indicating that the network is jammed.
\begin{figure}[tb]
\begin{center}
\includegraphics[width=8.2cm]{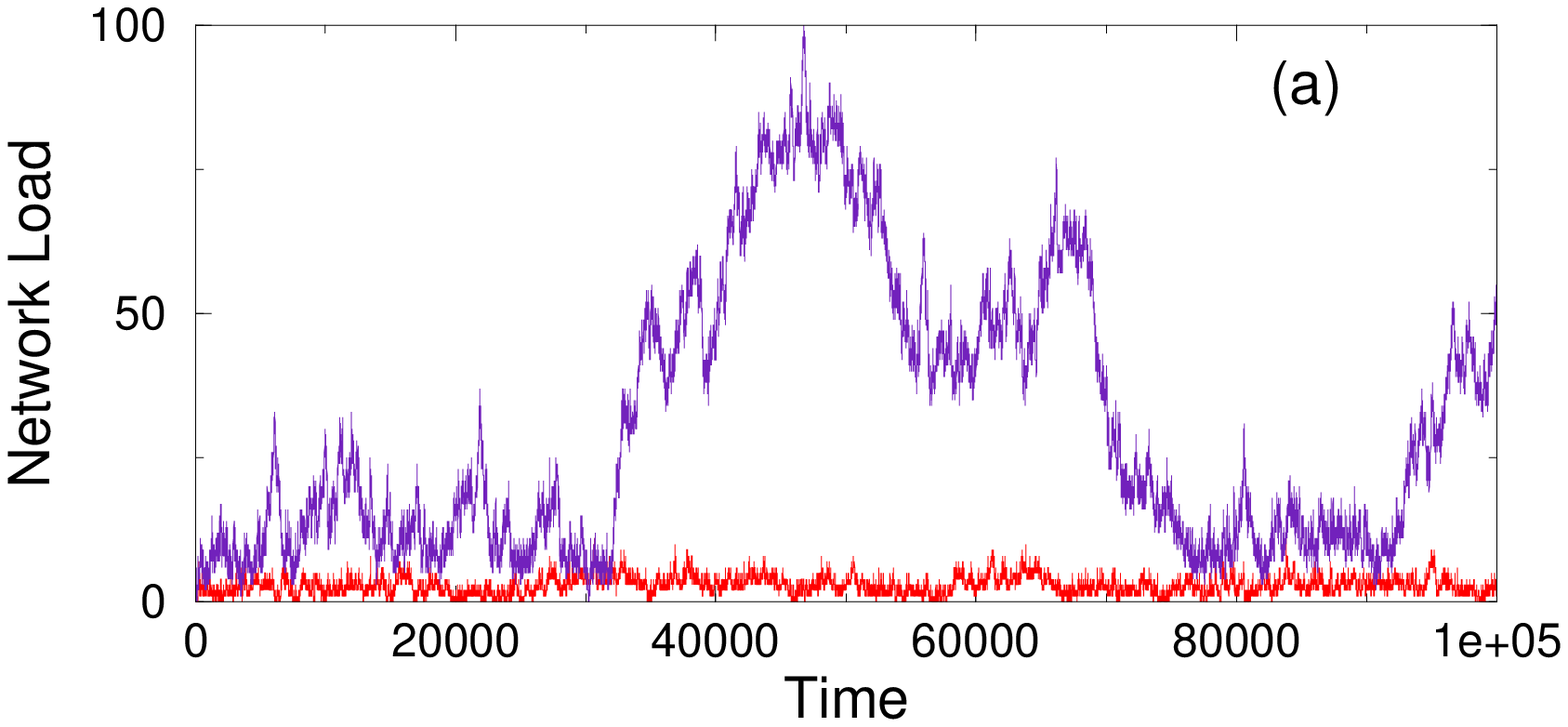}\\
\includegraphics[width=8.2cm]{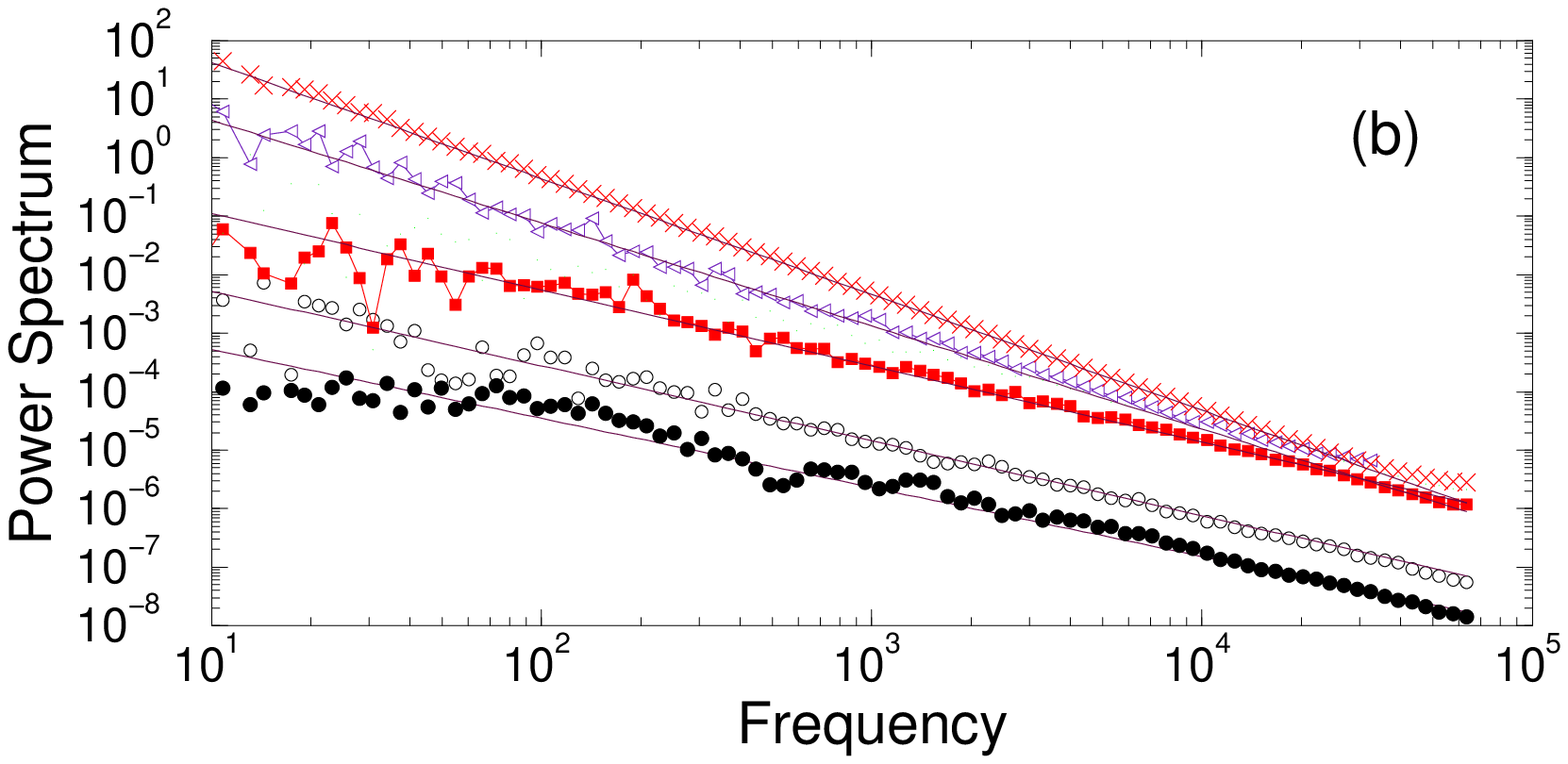}
\end{center}
\caption{(a) Network-load timeseries in transport on the Web-graph
for two different posting rates  
$R$ representing the stationary free flow ($R$=0.1, bottom line) 
and flow with a temporary crisis ($R$=0.3, top line).
(b) Power spectrum of the load timeseries for 
$R$=0.005, 0.1, 0.2, 0.3, and 0.4 (bottom to top). 
Data are log-binned. Fit lines have slopes $\phi =$ 1.18, 1.20, 
1.30, 1.76, and 1.98. Errors are within $\pm$0.02.}
\label{figgraph}
\end{figure} 
In Fig.\ 2 (a) we show an example of the network-load timeseries for two 
different posting rates.
For low posting rates queuing effects are small and 
transit-times are generally short. Consequently, the total number of 
particles 
on the network fluctuates around a small average value. 
For a much larger posting rate the flow is still stationary 
but the average number of particles is also much higher. 
In addition, the character of the fluctuations changes, with occasional
dramatic increases in the load, which then dissipate over a 
relatively long time-period. 
We considered very long timeseries in order to verify that the particle flow 
which contains these temporary crises is in fact stationary.
When the posting rate is increased over a certain value 
($R\approx 0.4$ in this particular case) the 
network-load exhibits a systematic increase, signaling 
jamming in the network.

The power spectra of the network-load timeseries are shown in 
Fig.\ 2 (b) for several values of $R$. 
When posting rates are such that stationary flow 
(with or without crises) occurs, the
work-load timeseries exhibit {\it anti-persistence}, 
with the exponent $\phi$ 
of the power-spectrum $S(f) \sim f^{-\phi }$ increasing with $R$ 
from $\phi=1.2$ towards $\phi=2$. 
This anti-persistence in the low $R$ regime may be 
attributed to the regulatory role of the super-structure
which is associated with the two hub types in the network. When the 
particle density is high, queues are formed at these nodes 
and the queuing particles, although still in transit,
are not contributing to the randomness of the load. 
Moreover, when a particle queuing at
a hub gets its turn to move, it gets quickly to its destination, 
which is often found in the next-nearest-neighborhood surrounding 
one of the hubs, and is then removed from the network.   
Another situation appears when the queue at a hub node is full. 
Then the dynamics
is reduced to {\it one particle out, one particle in}, which entirely destroys 
the correlations in the transport. 
In the jammed flow regime  the temporal correlations in the network-load are 
entirely lost (reflected $\phi \approx 2$ within the error bars).

Thus the main properties of the traffic (and jamming)  
are related to the hubs and their associated structure. 
To further investigate this effect, we have also studied  
how the activity is distributed over the network. 
In Fig.\ 3 we show the power spectrum of the 
number of active nodes (non-empty nodes) for varying posting rates $R$. 
\begin{figure}[tb]
\begin{center}
\includegraphics[width=8.2cm]{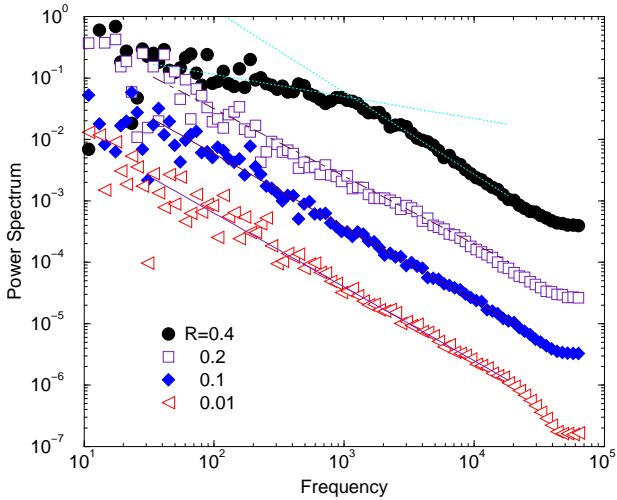}
\end{center}
\caption{Power spectrum of network's activity 
(number of simultaneously active nodes) for different 
values of posting rates $R$. Data are log-binned and the upper 
two curves shifted vertically for clarity of the plot. 
Full lines have slopes -1.2, dashed-dotted line: -1.1, 
and two dotted lines are indicating two slopes of 
-0.36 and 1.36.}
\label{figgraph}
\end{figure} 
In the free flow 
regime at low posting rate the timeseries of the network's activity 
is also {\it anti-persistent} with a well defined slope in the power 
spectrum. For intermediate posting rates,
where particle density is increased and temporary crises 
occur in the flow, a new gradient
starts to develop  in the low frequency range. 
Eventually, at high particle density,
two distinct types of behavior occur for 
the low and for high frequencies, indicating that 
a part of the network has entered the jammed regime.

\section{Waiting-times and network-load distributions}

The dependence on posting rates of the temporal fluctuations in the 
network-load and activity,  as observed in the previous section,  
is also found in the averaged statistical properties.
The probability density function of the overall network-load,
shown in Fig.\ 4,  is obtained from a histogram from 
timeseries like those in Fig.\ 2 and averaging over a 
number of network configurations.

At very low traffic density queuing occurs rarely.
For intermediate densities queuing becomes more important and consequently
the load distribution gets a tail. The large overall load 
appears -- although with smaller probability -- to be 
related to the volatile fluctuations
discussed above. In addition, the dominant part of the distribution 
becomes of log-normal type with the mean shifted towards 
higher values with increased $R$.
On approaching the jamming limit the tail of the distribution 
becomes more pronounced, suggesting 
that queues and thus waiting times increase in a large part of the network.
\begin{figure}[tb]
\begin{center} 
\includegraphics[width=8.2cm]{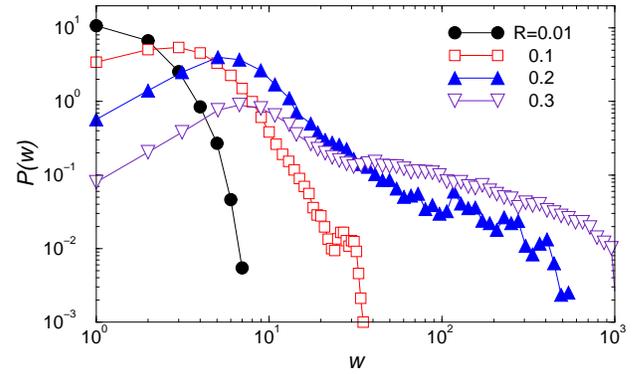}
\end{center}
\caption{Probability density function of the network-load 
in the stationary regime with free flow and flow with temporary 
crises. Data averaged over 100 network realizations and is log-binned.}
\label{figgraph}
\end{figure} 
The distribution of waiting-times of individual particles shows substantial
changes when temporary jamming in the network starts 
occurring more frequently. 
In Fig.\ 5 we show the distribution of waiting-times when 
the posting rate is varied
from low to intermediate $R$, where the flow is still stationary.

\begin{figure}[tb]
\begin{center} 
\includegraphics[width=8.2cm]{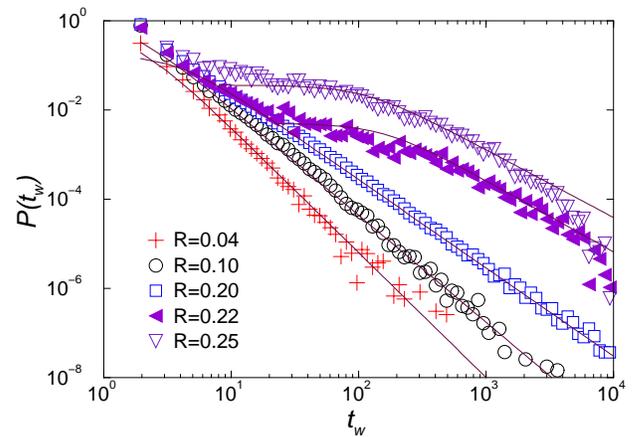}
\end{center}
\caption{Distribution of waiting-times of individual particles for varying 
posting rates $R$ in the free flow and flow with temporary crisis. Data 
collected from a particle which moves within a computation 
time-window of 200000 time-steps and is log-binned. }
\label{figgraph}
\end{figure}

The distribution is
taken from particles which move within a total time-window 
of up to 200000 time-steps.
When crises start to occur the waiting-time 
distribution changes from simple
power-law to  a generalized Cauchy-type distribution 
with a slope varying again with the posting rate $R$. 
There is robustness in the system in the sense that minor changes to the 
queuing discipline or the buffer sizes do not lead to a
qualitative difference in these results.
However, if  the queuing discipline is changed from LIFO to
FIFO (first in first out) then the power dependence of the 
waiting-time distribution  disappears. 
In particular, for the range of posting rates below jamming, for FIFO-queuing 
a gap in the waiting-time distributions appears, separating the
power-law like tail for long waiting-times at hubs, from the short 
waiting-times, which are found at most of the other nodes in the network. 
In LIFO-queuing the waiting-time  at a given node is dynamically 
conditioned by incoming packet streams from all neighboring nodes 
{\it after} arrival of the packet. In contrast, much weaker 
dependence on the local
network structure is incorporated in the FIFO-queuing mechanism, where  
the waiting-time is exactly given by the queue length of that node 
{\it at arrival} of the packet, independently on how long it took 
the network to build  the queue \cite{TT}.  
However, the transit-time distribution, which integrates waiting-times 
over many nodes along the packet trajectory on the network, remains a 
power-law for both queuing mechanisms at low and moderate $R$ \cite{TT,TR}.

\section{Conclusions}

We have performed an extensive study of
both the microscopic dynamics (timeseries) and macroscopic
 probability density functions of network-load and waiting-times of particles 
in a model of transport on Web-graphs. Particles move using a 
local search algorithm with next-nearest-neighbor signaling, 
which uses the underlying network topology efficiently.
We have demonstrated how  network function changes when
posting rates are altered. 

Statistical properties of both the microscopic dynamics and the 
probability density functions suggest that there occur three flow regimes, 
depending on the overall traffic density:
stationary free flow at low posting rates; 
stationary flow with temporary jams which are subsequently 
slowly dissipated by the system, and jammed flow at high posting rates.  
Our  analysis applies to the stationary free flow and flow with
temporary crisis, whereas we can recognize the approach to jamming  
transition from the low-density side. In particular, we find that 
the jamming threshold 
is marked by the loss of temporal correlations in the work-load timeseries
and by permanent increase of the overall network load. 
The waiting-times of packets diverge on approaching the jammed flow, therefore 
the statistical analysis of waiting-times can not be carried out in the 
vicinity of the transition.

The super-structure associated with the two types of hubs in the network 
plays an essential role in determining the properties of the traffic. 
In the low particle density regime it contributes to
efficient free flow and the clearance of temporary jamming 
through self-regulating
mechanisms which may be directly measured 
by the degree of anti-persistence in the work-load timeseries. 
However, at high particle density the hubs are the first nodes to jam,
forcing a crucial  part of the network 
structure to enter a slow-traffic (jammed) 
regime, whereas the rest of the network, which carries 
much less traffic, may continue to function normally.
The hub with higher connectivity is likely to jam first. Then the network
continues to function with the other hub until it also jams, eventually 
causing the congestion to spread over the associate structure.  
In principle, the relative distance between hubs plays the role in spreading
of the congestion, e.g., when hubs are far apart the jamming in two
parts of the network will occur almost independently. In our model the 
two-hub structure is an emergent feature and the distance between hubs
can not be controlled by the network growth rules in our present setting. 
However, due to strong clustering property, the relative 
distance between the hubs is rather short and -- in most network 
realizations -- within the reach of the next-nearest-neighbor search.   
Thus highly organized scale-free 
networks, containing these super-structures, operate more efficiently than 
conventional networks for a wide range of driving conditions. 
However, their advantage may become a weak point when the conditions
change, i.e., when the traffic density increases over a certain limit. 
This property of traffic may also be important to prevent  dynamical 
attacks  which target hubs, such as denial of service attacks 
to highly connected servers.
Our overall conclusion is, 
although the particular topology of the Web-graph is essential for its 
efficient operation under normal conditions, the same topology may also  
be a weakness under different conditions, reflected by vanishing 
anti-correlations in network-load timeseries. 
It seems clear that to assess the 
vulnerability of a network to different types of attack, one cannot just 
consider the topology of a network, one must consider how that topology
influences particle transport on it.


\end{document}